\documentclass[twocolumn,english,pra]{revtex4-1}
\usepackage[T1]{fontenc}
\usepackage[latin9]{inputenc}
\usepackage{color}
\usepackage{amsmath}
\usepackage{amssymb}
\usepackage{graphicx}
\usepackage{esint}

\makeatletter
\@ifundefined{textcolor}{}
{%
 \definecolor{BLACK}{gray}{0}
 \definecolor{WHITE}{gray}{1}
 \definecolor{RED}{rgb}{1,0,0}
 \definecolor{GREEN}{rgb}{0,1,0}
 \definecolor{BLUE}{rgb}{0,0,1}
 \definecolor{CYAN}{cmyk}{1,0,0,0}
 \definecolor{MAGENTA}{cmyk}{0,1,0,0}
 \definecolor{YELLOW}{cmyk}{0,0,1,0}
 }

\@ifundefined{definecolor}
 {\usepackage{color}}{}
\makeatother

\makeatother

\usepackage{babel}
\begin{document}

\title{Entanglement and nonlocality in multi-particle systems }

\author{M. D. Reid, Q. Y. He and P. D. Drummond}

\affiliation{ARC Centre of Excellence for Quantum-Atom Optics, Centre for Atom
Optics and Ultrafast Spectroscopy, Swinburne University of Technology,
Melbourne 3122, Australia}
\begin{abstract}
\textbf{Entanglement, the Einstein-Podolsky-Rosen (EPR) paradox and
Bell's failure of local-hidden-variable (LHV) theories are three historically
famous forms of {}``quantum nonlocality''. We give experimental
criteria for these three forms of nonlocality in multi-particle systems,
with the aim of better understanding the transition from microscopic
to macroscopic nonlocality. We examine the nonlocality of $N$ separated
spin $J$ systems. First, we obtain multipartite Bell inequalities
that address the correlation between spin values measured at each
site, and then we review spin squeezing inequalities that address
the degree of reduction in the variance of collective spins. The latter
have been particularly useful as a tool for investigating entanglement
in Bose-Einstein condensates (BEC). We present solutions for two topical
quantum states: multi-qubit Greenberger-Horne-Zeilinger (GHZ) states,
and the ground state of a two-well BEC. }


\end{abstract}
\maketitle

\section{\textbf{Introduction}}

Nonlocality in quantum mechanics has been extensively experimentally
investigated. Results to date support the quantum prediction, first
presented by Bell, that quantum theory is inconsistent with a combination
of premises now generally called {}``local realism'' \cite{Bell,CHSH}. 

However, the extent that quantum mechanics is inconsistent with local
realism at a more mesoscopic or macroscopic level is still not well
understood. Schr\"odinger presented the case that loss of realism
macroscopically would be a concern, and raised the question of how
to link the loss of local realism with macroscopic superposition states
\cite{Schrodinger-1,Schrodinger-2,Schrodinger-3,legg}.

The advent of entangled Bose-Einstein condensate (BEC) states leads
to new possibilities for testing mesoscopic and macroscopic quantum
mechanics. With this in mind, the objective of this article is to
give an overview of a body of work that explores nonlocality in multi-particle
or multi-site systems. 

Three types of nonlocality are reviewed: \emph{entanglement }\cite{Schrodinger-1},
the \emph{Einstein-Podolsky-Rosen (EPR) paradox} \cite{epr}, and\emph{
Bell's nonlocality} \cite{Bell,CHSH}. Examples of criteria to demonstrate
each of these nonlocalities is presented, first for multi-site {}``qubits''
(many spin $1/2$ particles) and then for multi-site {}``qudits''
(many systems of higher dimensionality such as high spin particles). 

The criteria presented in this paper are useful for detecting the
nonlocality of the $N$-qubit (or $N$-qudit) Greenberger-Horne-Zeilinger
(GHZ) states \cite{ghz,mermin90}. These states are extreme superpositions
that were shown by GHZ to demonstrate a very striking {}``all or
nothing'' type of nonlocality. This nonlocality can manifest as a
violation of a Bell inequality, and at first glance these violations,
because they increase exponentially with $N$, appear to indicate
a more extreme nonlocality as the size $N$ of the system increases
\cite{mermin bellghz}. 

We point out, however, that the detection of \emph{genuine} $N$-body
nonlocality, as first discussed by Svetlichny \cite{genuine,collspinmol},
requires much higher thresholds. Genuine $n$-party nonlocality (e.g.
genuine entanglement) requires that the nonlocality is shared among
\emph{all} $N$ parties, or particles. The violations in this case
do not increase with $N$, and the detection over many sites is very
sensitive to loss and inefficiencies.

Finally, we review and outline how to detect entanglement \cite{collspinmol}
and the EPR paradox using collective spin measurements. This approach
has recently been employed to establish a genuine entanglement of
many particles in a BEC \cite{exp multi,treutnature}.

\section{Three Famous types of nonlocality}

The earliest studies of nonlocality concerned bipartite systems. Einstein-Podolsky-Rosen
(EPR) \cite{epr} began the debate about quantum nonlocality, by pointing
out that for some quantum states there exists an inconsistency between
the premises we now call {}``local realism'' and the completeness
of quantum mechanics.

\emph{Local realism} (LR) may be summarized as follows. EPR argued
\cite{epr,epr rev } first for {}``locality'', by claiming that
there could be no {}``action-at-a-distance''. A measurement made
at one location cannot instantaneously affect the outcomes of measurements
made at another distant location. EPR also argued for {}``reality'',
which they considered in the following context. Suppose one can predict
with certainty the result of a measurement made on a system, without
disturbing that system. Realism implies that this prediction is possible,
only because the outcome for that measurement was a predetermined
property of the system. EPR called this predetermined property an
{}``\emph{element of reality}'', though most often the element of
reality is interpreted as a {}``\emph{hidden variable}''. The essence
of EPR's local realism assumption is that results of measurements
made on a system at one location come about because of predetermined
properties of that system, and because of their local interactions
with the measurement apparatus, not because of measurements that are
made simultaneously at a distant locations.

\subsection{EPR paradox}

EPR argued that for states such as the spin $1/2$ singlet state 
\begin{equation}
|\psi\rangle=\frac{1}{\sqrt{2}}\left(|\uparrow\rangle_{A}|\downarrow\rangle_{B}-|\downarrow\rangle_{A}|\uparrow\rangle_{B}\right)\label{eq:spinbohm}
\end{equation}
there arises an inconsistency of the LR premises with the quantum
predictions. Here, we define $|\uparrow\rangle_{A/B}$ and $|\downarrow\rangle_{A/B}$
as the spin {}``up'' and {}``down'' eigenstates of $J_{A/B}^{Z}$
for a system at location $A/B$. For the state (\ref{eq:spinbohm}),
the prediction of the spin component $J_{A}^{Z}$ can be made by measurement
of the component $J_{B}^{Z}$ at $B$. From quantum theory, the two
measurements are perfectly anticorrelated. According to EPR's Local
Realism premise (as explained above), there must exist an {}``element
of reality'' to describe the predetermined nature of the spin at
$A$. We let this element of reality be symbolized by the variable
$\lambda_{z}$, and we note that $\lambda_{z}$ assumes the values
$\pm1/2$ (\ref{eq:spinbohm}).

Calculation shows that there is a similar prediction of a perfect
anti-correlation for the other spin component pairs. Therefore, according
to LR, each of the spin components $J_{A}^{Y}$ and $J_{A}^{X}$ can
also be represented by an element of reality, which we denote $\lambda_{x}$
and $\lambda_{y}$ respectively. A moment's thought tells us that
the if there is a state for which all three spins are completely and
precisely predetermined in this way, then this {}``state'' cannot
be a quantum state. Such a {}``state'' is generally called a {}``local
hidden variable (LHV) state'', and the set of three variables are
{}``hidden'', since they are not part of standard quantum theory.
Hence, EPR argued, quantum mechanics is incomplete.

Since perfect anticorrelation is experimentally impossible, an operational
criterion for an EPR paradox can be formulated as follows. Consider
two observables $X$ and $P$, with commutators like position and
momentum. The Heisenberg Uncertainty Principle is $\Delta X\Delta P\geq1$,
where $\Delta X$ and $\Delta P$ are the variances of the outcomes
of measurements for $X$ and $P$ respectively. The EPR paradox criterion
is \cite{reidepr} 
\begin{equation}
\Delta_{inf}X\Delta_{inf}P<1,\label{eq:eprcrit}
\end{equation}
where $\Delta_{inf}X\equiv V(X|O_{B})$ is the {}``variance of inference''
i.e. the variance of $X$ conditional on the measurement of an observable
$O_{B}$ at a distant location $B$. The $\Delta_{inf}P\equiv V(P|Q_{B})$
is defined similarly where $Q_{B}$ is a second observable for measurement
at $B$. This criterion reflects that the combined uncertainty of
inference is reduced below the Heisenberg limit. Of course, the reduced
uncertainty applies over an ensemble of measurements, where only one
of the conjugate measurements is made at a time. This criterion is
also applicable to optical quadrature observables, where it has been
experimentally violated, although without causal separation. With
spin commutators, other types of uncertainty principle can be used
to obtain analogous inferred uncertainty limits.

The demonstration of an EPR paradox through the measurement of correlations
satisfying Eq. (\ref{eq:eprcrit}) is a proof that local realism is
inconsistent with the completeness of quantum mechanics (QM). Logically,
one must: discard local realism, the completeness of QM, or both.
However, it does not indicate which alternative is correct.

\subsection{Schr\"odinger's Entanglement}

Schr\"odinger \cite{Schrodinger-1,Schrodinger-2,Schrodinger-3} noted
that the state (\ref{eq:spinbohm}) is a special sort of state, which
he called an an \emph{entangled} state. An entangled state is one
which cannot be factorized: for a pure state, we say there is entanglement
between $A$ and $B$ if we cannot write the composite state $|\psi\rangle$
(that describes all measurements at the two locations) in the form
$|\psi\rangle=|\psi\rangle_{A}|\psi\rangle_{B}$, where $|\psi\rangle_{A/B}$
is a state for the system at $A/B$ only. 

For mixed states, there is said to be entanglement when the density
operator for the composite system cannot be written as a mixture of
factorizable states \cite{peres}. A mixture of factorizable states
is said to be a \emph{separable} state, which where there are just
two sites, is written as
\begin{equation}
\rho=\sum_{R}P_{R}\rho_{A}^{R}\rho_{B}^{R}.\label{eq:sep2}
\end{equation}
If the density operator cannot be written as (\ref{eq:sep2}), then
the mixed system possesses \emph{entanglement} (between $A$ and $B$).
More generally, for $N$ sites, full separability implies
\begin{equation}
\rho=\sum_{R}P_{R}\rho_{1}^{R}...\rho_{N}^{R}.\label{eq:sepN}
\end{equation}
If the density operator cannot be expressed in the fully separable
form (\ref{eq:sepN}), there is entanglement between at least two
of the sites. 

We consider measurements $\hat{X}_{k}$, with associated outcomes
$X_{k}$, that can be performed on the $k$-th system ($k=1,...,N$).
For a separable state (\ref{eq:sepN}), it follows that the joint
probability for outcomes is expressible as
\begin{equation}
P(X_{1},...,X_{N})=\int_{\lambda}P(\lambda)P_{Q}(X_{1}|\lambda)...P_{Q}(X_{N}|\lambda)d\lambda\,,\label{eq:sepent}
\end{equation}
where we have replaced for convenience of notation the index $R$
by $\lambda$, and used a continuous summation symbolically, rather
than a discrete one, so that $P(\lambda)\equiv P_{R}$. The subscript
$Q$ represents {}``quantum'', because there exists the quantum
density operator $\rho_{k}^{\lambda}\equiv\rho_{k}^{R}$ for which
$P(X_{k}|\lambda)\equiv\langle X_{k}|\rho_{k}^{\lambda}|X_{k}\rangle$.
In this case, we write $P(X_{k}|\lambda)\equiv P_{Q}(X_{k}|\lambda)$,
where the subscript $Q$ reminds us that this is a quantum probability
distribution. The model (\ref{eq:sepent}) implies (\ref{eq:sep2})
\cite{wisesteer,wisesteer2}, and has been studied in Ref. \cite{eric steer},
in which it is referred to as a \emph{quantum separable model }(QS). 

We can test nonlocality when each system $k$ is spatially separated.
We will see from the next section that LR implies the form (\ref{eq:sepent}),
but without the subscripts {}``Q'', that is, without the underlying
local states designated by $\lambda$ necessarily being quantum states.
If the quantum separable QS model can be shown to fail where each
$k$ is spatially separated, one can only have consistency with Local
Realism if there exist underlying local states that are \emph{non-quantum.
}This is an EPR paradox, since it is an argument to complete quantum
mechanics, based on a requirement that LR be valid.

The EPR paradox necessarily requires entanglement \cite{epr rev ,mallon}.
The reason for this is that for separable states (\ref{eq:sep2}-\ref{eq:sepent}),
the uncertainty relation that applies to each of the states $|\psi\rangle_{A}$
and $|\psi\rangle_{B}$ will imply a minimum level of local uncertainty,
which means that the noncommuting observables cannot be sufficiently
correlated to obtain an EPR paradox. In other words, the entangled
state (\ref{eq:spinbohm}) can possess a greater correlation than
possible for (\ref{eq:sep2}). 

Schr\"odinger also pointed to two paradoxes \cite{Schrodinger-1,Schrodinger-2,Schrodinger-3}
in relation to the EPR paper. These gedanken-experiments strengthen
the apparent need for the existence of EPR {}``elements of reality'',
in situations involving macroscopic systems, or spatially separated
ones. The first is famously known as the Schr\"odinger's cat paradox,
and emphasizes the importance of EPR's {}``elements of reality''
at a \emph{macroscopic} level. Reality applied to the state of a cat
would imply a cat to be either dead or alive, prior to any measurement
that might be made to determine its {}``state of living or death''.
We can define an {}``element of reality'' $\lambda_{cat}$ , to
represent that the cat is \emph{predetermined} to be dead (in which
case $\lambda_{cat}=-1$) or alive (in which case $\lambda_{cat}=+1$).
Thus, the observer looking inside a box, to make a measurement that
gives the outcome {}``dead'' or {}``alive'', is simply uncovering
the value of $\lambda_{cat}$. Schr\"odinger's point was that the
element of reality specification is not present in the quantum description
$|\Psi\rangle=\frac{1}{\sqrt{2}}\left(|dead\rangle+|alive\rangle\right)$
of a superposition of two macroscopically distinguishable states. 

The second paradox raised by Schr\"odinger concerns the apparent
{}``action at-a-distance'' that seems to occur for the EPR entangled
state. Unless one identifies an element of reality for the outcome
$A$, it would seem to be the action of the measurement of $B$ that
immediately enables prediction of the outcome for the measurement
at $A$. Schr\"odinger thus introduced the notion of {}``\emph{steering}''.

While all these paradoxes require entanglement, we emphasize that
entanglement \emph{per se} is a relatively common situation in quantum
mechanics. It is necessary for a quantum paradox, but does not by
itself demonstrate any paradox.

\subsection{Bell's nonlocality: failure of local hidden variables (LHV)}

EPR claimed as a solution to their EPR paradox that hidden variables
consistent with local realism would exist to further specify the quantum
state. It is the famous work of Bell that proved the impossibility
of finding such a theory. This narrows down the two alternatives possible
from a demonstration of the EPR paradox, and shows that local realism
itself is invalid. 

Specifically, Bell considered the predictions of a Local Hidden Variable
(LHV) theory, to show that they would be different to the predictions
of the spin-half EPR state (\ref{eq:spinbohm}). Following Bell \cite{Bell,CHSH},
we have a \emph{local hidden variable model} (LHV) if the joint probability
for outcomes of simultaneous measurements performed on the $N$ spatially
separated systems is given by
\begin{equation}
P(X_{1},...,X_{N})=\int_{\lambda}P(\lambda)P(X_{1}|\lambda)...P(X_{N}|\lambda)d\lambda\,.\label{eq:bell}
\end{equation}
Here $\lambda$ are the {}``local hidden variables'' and $P(X_{k}|\lambda)$
is the probability of $X_{k}$ given the values of $\lambda$, with
$P(\lambda)$ being the probability distribution for $\lambda$. The
factorization in the integrand is Bell's locality assumption, that
$P(X_{k}|\lambda)$ depends on the parameters $\lambda$, and the
measurement choice made at $k$ only. The hidden variables $\lambda$
describe a\emph{ }local state\emph{ }for each site, in that the probability
distribution $P(X_{k}|\lambda)$ for the measurement at $k$ is given
as a function of the $\lambda$. The form of (\ref{eq:bell}) is formally
similar to (\ref{eq:sepN}) except in the latter there is the additional
requirement that the local states are quantum states. If (\ref{eq:bell})
fails, then we have\emph{ }proved\emph{ }a\emph{ }failure of all LHV
theories\emph{, }which we refer to as a \emph{Bell violation }or\emph{
Bell nonlocality} \cite{eric steer}\emph{. }

The famous Bell-Clauser-Horne-Shimony-Holt (CHSH) inequalities follow
from the LHV model, in the $N=2$ case. Bell considered measurements
of the spin components $J_{A}^{\theta}=\cos\theta J_{A}^{X}+\sin\theta J_{A}^{Y}$
and $J_{B}^{\theta}=\cos\phi J_{B}^{X}+\sin\phi J_{B}^{Y}$. He then
defined the spin product $E(\theta,\phi)=\langle J_{A}^{\theta}J_{B}^{\phi}\rangle$
and showed that for the LHV model, there is always the constraint
\begin{equation}
B=E(\theta,\phi)-E(\theta,\phi')+E(\theta',\phi)+E(\theta',\phi')\leq2.\label{eq:bellchsh}
\end{equation}
The quantum prediction for an entangled Bell state (\ref{eq:spinbohm})
is $E(\theta,\phi)=\cos(\theta-\phi)$ and the inequality is violated
for the choice of angles 
\begin{equation}
\theta=0,\theta'=\pi/2,\phi=\pi/4,\phi'=3\pi/4\label{eq:angles}
\end{equation}
for which the quantum prediction becomes $B=2\sqrt{2}$. Tsirelson's
theorem proves the value of $B=2\sqrt{2}$ to be the maximum violation
possible for any quantum state \cite{tsirel}. We note that experimental
inefficiencies mean that violation of the CHSH inequalities for causally
separated detectors is difficult, and has so far always required additional
assumptions in the interpretation of experimental data.

\subsection{Steering as a special nonlocality }

Recently, Wiseman et al (WJD) \cite{wisesteer,wisesteer2} have constructed
a hybrid separability model, called the Local Hidden State Model (LHS),
the violation of which is confirmation of Schr\"odinger's {}``steering''
(Figure 1). The bipartite local hidden state model (LHS) assumes 
\begin{equation}
P(X_{A},X_{B})=\int_{\lambda}P(\lambda)P(X_{A}|\lambda)P_{Q}(X_{B}|\lambda)d\lambda\,.\label{eq:bell-1}
\end{equation}
Thus, for one site $A$ which we call {}``Alice'', we assume a local
hidden variable (LHV) state, but at the second site $B$, which we
call {}``Bob'', we assume a local quantum state (LQS). The violation
of this model occurs iff there is a {}``\emph{steering}'' of Bob's
state by Alice \cite{steerexp}. 

WJD pointed out the association of steering with the EPR paradox \cite{wisesteer}.
The EPR criterion is also a criterion for steering, as defined by
the violation of the LHS model. An analysis of the EPR argument when
generalized to allow for imperfect correlation and arbitrary measurements
reveals that violation of the LHS model occurs iff there is an EPR
paradox \cite{eric steer,epr rev }. As a consequence, the violation
of the LHS model is referred to as demonstration of a type of nonlocality
called {}``\emph{EPR steering}'' \cite{eric steer}. EPR steering
confirms the incompatibility of local realism with the \emph{completeness}
of quantum mechanics, just as with the approach of EPR in their original
paper \cite{epr}.

The notion of steering can be generalized to consider $N$ sites,
or observers \cite{ericmulti}. The multipartite LHS model is (Figure
1)
\begin{multline}
P(X_{1},...,X_{N})=\\
\int d\lambda P(\lambda)\prod_{j=1}^{T}P_{Q}(X_{j}|\lambda)\prod_{j=T+1}^{N}P(X_{j}|\lambda),\label{eq:LHS_model_multipartite}
\end{multline}
where here we have $T$ quantum states, and $N-T$ LHV local states.
We use the symbol $T$ to represent the quantum states, since these
are the{}``trusted sites'' in the picture put forward by WJD \cite{wisesteer}.
This refers to an application of this generalized steering to a type
of quantum cryptography in which an encrypted secret is being shared
between sites. At some of the sites, the equipment and the observers
are trusted, while at other sites this is not the case.

In this picture, which is an application of the LHS model, an observer
$C$ wishes to establish entanglement between two observers Alice
and Bob. The violation of the QS model is sufficient to do this, provided
each of the two observers Alice and Bob can be trusted to report the
values for their local measurements. It is conceivable however that
they report instead statistics that can give a violation of LHS model,
so it seems as if there is entanglement when there is not. WJD point
out the extra security present if instead there is the stronger requirement
of violation of the LHV model, in which the untrusted observers are
identified with a LHV state. 

Cavalcanti et al \cite{ericmulti} have considered the multipartite
model (\ref{eq:LHS_model_multipartite}), and shown that violation
of (\ref{eq:LHS_model_multipartite}) where $T=1$ is sufficient to
imply an \emph{EPR steering} paradox exists between at least two of
the sites. Violation where $T=0$ is proof of Bell's nonlocality,
and violation where $T=N$ is a confirmation of entanglement (quantum
inseparability).

\subsection{Hierarchy of nonlocality }

WJD established formally the concept of a hierarchy of nonlocality
\cite{wisesteer,wisesteer2}. Werner \cite{werner} showed that some
classes of entangled state can be described by Local Hidden Variable
theories and hence cannot exhibit a Bell nonlocality. WJD showed that
not all entangled states are {}``steerable'' states, defined as
those that can exhibit EPR steering. Similarly, they also showed that
not all EPR steerable states exhibit Bell nonlocality. However, we
see from the definitions that all EPR steering states must be entangled,
and all Bell-nonlocal states ( defined as those exhibiting Bell nonlocality)
must be EPR steering states. Thus, the Bell-nonlocal states are a
strict subset of EPR steering states, which are a strict subset of
entangled states, and a hierarchy of nonlocality is established.

\begin{figure}[h]
\begin{centering}
\includegraphics[scale=0.35]{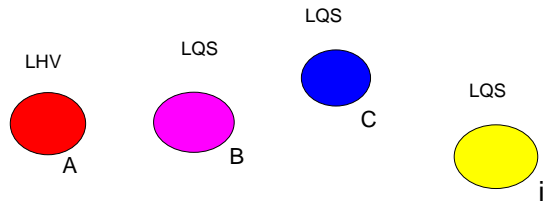}\medskip{}

\par\end{centering}
\caption{\emph{The LHS model:} Multiple {}``local'' systems at are different
spatial locations; the separability model is assumed. Some local systems
are constrained only to be described by LHV, while others are constrained
to be quantum systems (LQS).}
\end{figure}

\section{Multiparticle Nonlocality}

Experiments that have been performed on many microscopic systems support
quantum mechanics. Those that test Bell's theorem \cite{Bell,CHSH},
or the equivalent, are the most useful, since they directly refute
the assumption of local realism. While these experiments still require
additional assumptions, it is generally expected that improved technology
will close the remaining loopholes.

There remains however the very important question of whether reality
will hold macroscopically. Quantum mechanics predicts the possibility
of superpositions of two macroscopically distinguishable states \cite{legg},
like a cat in a superposition of dead and alive states. Despite the
apparent paradox, there is increasing evidence for the existence of
mesoscopic and macroscopic quantum superpositions.

As with microscopic systems, there is a need to verify the loss of
reality for macroscopic superpositions in an objective sense, by following
Bell's example and comparing the predictions of quantum mechanics
with those based on premises of local realism. The first steps toward
this have been taken, through theoretical studies of nonlocality for
multi-particle systems. Two limits have been rather extensively examined.
The first is that of bipartite qudits. The second is multipartite
qubits. Surprisingly, while it may have been thought that the violation
of LR would diminish or vanish at a critical number of particles,
failure of local realism has been shown possible according to quantum
mechanics, for arbitrarily large numbers of particles. The third possibility
of multipartite qudits has not been treated in as much detail.

\subsection{Bipartite qudits}

The simplest mesoscopic extension of the Bell case (\ref{eq:spinbohm})
is to consider bipartite qudits: two sites of higher dimensionality.
The maximally entangled state in this case is 
\begin{equation}
|\psi\rangle=\frac{1}{\sqrt{d}}\sum_{j=0}^{d-1}|jj\rangle,\label{eq:maxentqudit}
\end{equation}
where $|jj\rangle$ is abbreviation for $|j\rangle_{A}|j\rangle_{B}$,
and $d$ is the dimensionality of the systems at $A$ and $B$. In
this case at each site $A$ and $B$ the possible outcomes are $j=0,...,d-1$.
This system can be realized by two spin $J$ systems, for which the
outcomes are $x$ given by $-J,-J+1,...,J-1,J$, so that $d=2J+1$,
and $j$ of Eq. (\ref{eq:maxentqudit}) is $j\equiv x+J$ where $x$
is the outcome of spin. It can also be realized by multi-particle
systems.

It was shown initially by Mermin, Garg and Drummond, and Peres and
others \cite{highd,multibell,drumspinbell,peresspin,gsisspin,franmrspin}
that quantum systems could violate local realism for large $d$. The
approach was to use the classic Bell inequalities derived for binary
outcomes. 

Later, Kaszlikowski et al \cite{high D K} showed that for maximally
entangled states (\ref{eq:maxentqudit}), the strength of violation
actually becomes stronger for increasing $d$. A new set of Bell inequalities
for bipartite qudits was presented by Collins et al (CGLMP) et al
\cite{collins high d,fuqdit} and it was shown subsequently by Acin
et al \cite{acin} that greater violations can be obtained with non-maximally
entangled states, and that the violations increase with $d$. Chen
et al \cite{chen} have shown that the violation of CGLMP inequalities
increases as $d\rightarrow\infty$ to a limiting value.

We wish to address the question of how the entanglement and EPR steering
nonlocalities increase with $d$. Since Bell nonlocality implies both
EPR steering and entanglement, these nonlocalities also increase with
$d.$ However, since there are distinct nested classes of nonlocality,
the violation could well be greater, for an appropriate set of measures
of the nonlocalities, and this problem is not completely solved for
the CGLMP approach. We later investigate alternative criteria that
show differing levels of violation for the different classes of nonlocality.

\subsection{Multipartite qubits: MABK Bell inequalities}

The next mesoscopic - macroscopic scenario that we will consider is
that of many distinct single particles $-$ the multi-site qubit system.
The interest here began with the Greenberger-Horne-Zeilinger (GHZ)
argument \cite{ghz}, which revealed a more extreme {}``all-or-nothing''
form of nonlocality for the case of three and four spin $1/2$ particle
(three or four qubits), prepared in a so-called GHZ state. The N qubit
GHZ state is written
\begin{equation}
|\Psi\rangle_{GHZ}=\frac{1}{\sqrt{2}}\{|0\rangle^{\otimes N}+|1\rangle^{\otimes N}\},\label{eq:ghz-1}
\end{equation}
where $|0\rangle$ and $|1\rangle$ in this case are spin up/ down
eigenstates. Mermin then showed that for this extreme superposition,
there corresponded a greater violation of LR, in the sense that the
new {}``Mermin'' Bell inequalities were violated by an amount that
increased exponentially with $N$ \cite{mermin bellghz}. These new
multipartite Bell inequalities of Mermin were later generalized by
Ardehali, Belinski and Klyshko, to give a set of MABK Bell inequalities
\cite{ard,bkmabk}.

The MABK inequalities define moments like $\langle J_{A}^{+}J_{B}^{+}J_{C}^{-}\rangle$,
where $J^{\pm}=J^{X}\pm iJ^{Y}$ and $J^{X}$, $J^{Y}$, $J^{Z}$,
$J^{2}$ are the standard quantum spin operators. In the MABK case
of qubits, Pauli operators are used, so that the spin outcomes $\pm1/2$
are normalized to $\pm1$. The $J^{X/Y}$ are redefined accordingly.
The moments are defined generally by 
\begin{equation}
\prod_{N}=\langle\Pi_{k=1}^{N}J_{k}^{s_{k}}\rangle\label{eq:prodNmabk}
\end{equation}
where $s_{k}=\pm1$ and $J^{s_{1}}\equiv J^{+}$ and $J^{s_{-1}}\equiv J^{-}$.
A LHV theory expresses such moments as the integral of a complex number
product: 
\begin{equation}
\prod_{N}=\int d\lambda P(\lambda)\Pi_{N,\lambda}\label{eq:prodNmabkhidden}
\end{equation}
where $\Pi_{N,\lambda}=\Pi_{k=1}^{N}\langle J_{k}^{s_{k}}\rangle_{\lambda}$
and $\langle J_{k}^{\pm}\rangle_{\lambda}=\langle J_{k}^{X}\rangle\pm i\langle J_{k}^{Y}\rangle$
where $\langle J_{k}^{X/Y}\rangle_{\lambda}$ is the expected value
of outcome for measurement $J^{X/Y}$ made at site $k$ given the
local hidden state $\lambda$. The $\Pi_{N,\lambda}$ is a complex
number product, which Mermin \cite{mermin bellghz} showed has the
following extremal values: for $N$ odd, a magnitude $2^{N/2}$ at
angle $\pi/4$ to real axis; for $N$ even, magnitude $2^{N/2}$ aligned
along the real or imaginary axis. With this algebraic constraint,
LR will imply the following inequalities, for odd $N$:
\begin{equation}
Re\prod_{N},\, Im\prod_{N}\leq2^{(N-1)/2}.\label{eq:mabkodd}
\end{equation}
For even $N$, the inequality $ $$Re\prod_{N},\, Im\prod_{N}\leq2^{N/2}$
will hold. However, it is also true, for even $N$, that \cite{ard}
\begin{equation}
Re\prod_{N}+Im\prod_{N}\leq2^{N/2}.\label{eq:mabkeven}
\end{equation}
The Eqns (\ref{eq:mabkodd}-\ref{eq:mabkeven}) are the MABK Bell
inequalities \cite{bkmabk}. Maximum violation of these inequalities
is obtained for the $N$-qubit Greenberger-Horne-Zeilinger (GHZ) state
(\ref{eq:ghz-1}) \cite{wernerwolf}. For optimal angle choice, a
maximum value 
\begin{equation}
\langle\mathrm{Re}\Pi_{N}\rangle,\mathrm{\langle Im}\Pi_{N}\rangle=2^{N-1}\label{eq:qm1}
\end{equation}
can be reached for the left -side of (\ref{eq:mabkodd}), while for
a different optimal angle choice, the maximum value 
\begin{equation}
\langle\mathrm{Re}\Pi_{N}\rangle+\langle\mathrm{Im}\Pi_{N}\rangle=2^{N-1/2}\label{eq:qm2}
\end{equation}
can be reached for the left-side of (\ref{eq:mabkeven}). MABK Bell
inequalities became famous for the prediction of exponential gain
in violation as the number of particles (sites), $N$, increases.
The size of violation is most easily measured as the ratio of left-side
to right-side of the inequalities (\ref{eq:mabkodd},\ref{eq:mabkeven}),
seen to be $2^{(N-1)/2}$ for the MABK inequalities. Werner and Wolf
\cite{wernerwolf} showed the quantum prediction to be maximum for
two-setting inequalities.

\subsection{MABK-type EPR steering and entanglement inequalities for multipartite
qubits}

Recently, MABK-type inequalities have been derived for EPR steering
and entanglement \cite{ericmulti}. Entanglement is a failure of quantum
separability, where each of the local states in (\ref{eq:LHS_model_multipartite})
are quantum states ($T=N$). EPR steering occurs when there is failure
of the LHS model with $T=1$. To summarize the approach of Ref. \cite{ericmulti},
we note the statistics of each \emph{quantum }state must satisfy a
quantum uncertainty relation
\begin{equation}
\Delta^{2}J^{X}+\Delta^{2}J^{Y}\geq1.\label{eq:hup-1}
\end{equation}
As a consequence, for every \emph{quantum} local state $\lambda$,
\begin{equation}
\langle J^{X}\rangle^{2}+\langle J^{Y}\rangle^{2}\leq1,\label{eq:hupconseqquantum}
\end{equation}
which implies the complex number product can have arbitrary phase,
leading to the new nonlocality inequalities, which apply for all $N$,
even or odd, and $T>0$:
\begin{eqnarray}
\langle\mathrm{Re}\Pi_{N}\rangle,\langle\mathrm{Im}\Pi_{N}\rangle & \leq & 2^{(N-T)/2},\label{eq:merminsteer}\\
\langle\mathrm{Re}\Pi_{N}\rangle+\langle\mathrm{Im}\Pi_{N}\rangle & \leq & 2^{(N-T+1)/2}.\label{eq:merminsteerstat-2}
\end{eqnarray}
For $T=N$, these inequalities if violated will imply entanglement,
as shown by Roy \cite{roy}. If violated for $T=1$, there is EPR
steering. As pointed out in \cite{ericmulti}, the exponential gain
factor of the violation with the number of particles $N$ increases
for increasing $T$: the strength of violation as measured by left
to right side ratio is $2^{(N+T-2)/2}$, but for both inequalities
(\ref{eq:merminsteer}-\ref{eq:merminsteerstat-2}).

\subsection{CFRD Multipartite qudit Bell, EPR steering and entanglement inequalities}

We now summarize an alternative approach to nonlocality inequalities,
developed by Cavalcanti, Foster, Reid and Drummond (CFRD) \cite{cfrd,vogelcfrd,acincfrd,cfrd he func,cfrdhepra}.
These hold for any operators, and are not restricted to spin-half
or qubits. We shall apply this approach to the case of a hierarchy
of inequalities, with some quantum and some classical hidden variable
states. Consider 
\begin{eqnarray}
|\prod_{N}| & \leq & \int d\lambda P(\lambda)\Pi_{k=1}^{N}|\langle J_{k}^{s_{k}}\rangle_{\lambda}|\label{eq:prodNmabkhidden-1}\\
 & = & \int d\lambda P(\lambda)\Pi_{k=1}^{N}\{\langle J_{k}^{X}\rangle_{\lambda}^{2}+\langle J_{k}^{Y}\rangle_{\lambda}^{2}\}^{1/2}.
\end{eqnarray}
We can see that for any LHV, because the variance is always positive,
one can derive an inequality for any operator 
\begin{equation}
\langle J_{k}^{X}\rangle_{\lambda}^{2}+\langle J_{k}^{Y}\rangle_{\lambda}^{2}\leq\langle(J_{k}^{X})^{2}\rangle_{\lambda}+\langle(J_{k}^{Y})^{2}\rangle_{\lambda}\label{eq:LHVspinvar}
\end{equation}
but then for a quantum state in view of the uncertainty relation (\ref{eq:hup-1}),
it is the case that for qubits (spin-1/2) 
\begin{equation}
\langle J_{k}^{X}\rangle_{\lambda}^{2}+\langle J_{k}^{Y}\rangle_{\lambda}^{2}\leq\langle(J_{k}^{X})^{2}\rangle_{\lambda}+\langle(J_{k}^{Y})^{2}\rangle_{\lambda}-1.\label{eq:lqsvarhur}
\end{equation}
For the particular case of qubits, the outcomes are $\pm1$ so that
simplification occurs, to give final bounds based on local realism
that are identical to (\ref{eq:merminsteer}-\ref{eq:merminsteerstat-2}).
We note that at $T=0$, there is also a CFRD Bell inequality, but
it is weaker than that of MABK, in the sense that the violation is
not as strong as is not predicted for $N=2$. Since this approach
holds for any operator, we now can generalize to arbitrary spin.

The expression (\ref{eq:LHVspinvar}-\ref{eq:lqsvarhur}) also holds
for arbitrary spin, for which case we revert to the usual spin outcomes
(rather than the Pauli spin outcomes of $\pm1$). The LHV result for
arbitrary spin is constrained by (\ref{eq:LHVspinvar}). The quantum
result however requires a more careful uncertainty relation that is
relevant to higher spins. In fact, for systems of fixed dimensionality
$d$, or fixed spin $J$, the {}``qudits'', the following uncertainty
relation holds 
\begin{equation}
\Delta^{2}J^{X}+\Delta^{2}J^{Y}\geq C_{J},\label{eq:cj}
\end{equation}
where the $C_{J}$ has been derived and presented in Ref. \cite{cj}.
The use of the more general result (\ref{eq:cj}) gives the following
higher-spin (qudit) nonlocality inequalities derived in Ref. \cite{higherspin steerq}:
\begin{multline}
|\langle\prod_{k=1}^{N}J_{k}^{s_{k}}\rangle|^{2}\leq\int d\lambda P(\lambda)\prod_{k=1}^{N}|\langle J_{k}^{s_{k}}\rangle_{\lambda}|^{2}\\
\leq\left\langle \prod_{k=1}^{T}(J_{k}^{X})^{2}+(J_{k}^{Y})^{2}-C_{k})\prod_{k=T+1}^{N}(J_{k}^{X})^{2}+(J_{k}^{Y})^{2}\right\rangle .\label{eq:ineqcom}
\end{multline}
Thus:
\begin{enumerate}
\item Entanglement is verified if ($T=N$)
\begin{eqnarray}
|\langle\prod_{k=1}^{N}J_{k}^{s_{k}}\rangle|^{2} & > & \langle\prod_{k=1}^{N}[(J_{k})^{2}-(J_{k}^{Z})^{2}-C_{J}]\rangle.\label{eq:spinjent}
\end{eqnarray}

\item An EPR-steering nonlocality is verified if ($T=1$) 
\begin{eqnarray}
|\langle\prod_{k=1}^{N}J_{k}^{s_{k}}\rangle|^{2} & > & \langle[(J_{1})^{2}-(J_{1}^{Z})^{2}-C_{J}]\nonumber \\
 &  & \ \ \times\prod_{k=2}^{N}[(J_{k}^{X})^{2}+(J_{k}^{Y})^{2}]\rangle.\label{eq:spinjsteer}
\end{eqnarray}

\item Bell inequality ($T=0$). The criterion to detect failure of the LHV
theories is 
\begin{eqnarray}
|\langle\prod_{k=1}^{N}J_{k}^{s_{k}}\rangle|^{2} & > & \langle\prod_{k=0}^{N}[(J_{k}^{X})^{2}+(J_{k}^{Y})^{2}]\rangle.
\end{eqnarray}

\end{enumerate}
These criteria will be called the {}``$C_{J}"$ CFRD nonlocality
criteria, and allow investigation of nonlocality in multisite qudits,
where the spin $J$ is fixed.

We investigate predictions for quantum states that are maximally entangled,
or not so\textcolor{black}{, according to measures of entanglement
that are justified for pure states. Maximally-entangled, highly correlated
states for a fixed spin $J$ are written
\begin{eqnarray}
|\Psi\rangle_{max} & = & \frac{1}{\sqrt{d}}\sum_{m=-J}^{J}|m\rangle_{1}|m\rangle_{2}....|m\rangle_{N}\nonumber \\
 & = & \frac{1}{\sqrt{d}}\sum_{j=0}^{d-1}|j\rangle_{1}|j\rangle_{2}...|j\rangle_{N},\label{eq:maxentst}
\end{eqnarray}
where $|m\rangle_{k}\equiv|J,m\rangle_{k}$ is the eigenstate of $J_{k}^{2}$
and $J_{k}^{Z}$ (eigenvalue $m$ for $J_{k}^{Z}$), defined at site
$k$, and the dimensionality is $d=2J+1$. This state is the extension
of (\ref{eq:maxentqudit}) for multiple sites. We follow \cite{acin}
however and consider more generally the non-maximally entangled but
highly correlated spin states of form
\begin{eqnarray}
|\psi\rangle_{non} & = & \frac{1}{\sqrt{n}}[r_{-J}|J,-J\rangle^{\otimes N}+r_{-J+1}|J,-J+1\rangle^{\otimes N}\nonumber \\
 &  & \ \ \ \ +...+r_{J}|J,+J\rangle^{\otimes N}],\label{eq:nonmaximally state}
\end{eqnarray}
where $|J,m\rangle^{\otimes N}=\Pi_{k=1}^{N}|J,m\rangle_{k}$, $n={\displaystyle \sum_{m=-J}^{J}}r_{m}^{2}$.
Here we will restrict to the case of real parameters symmetrically
distributed around $m=0$. The amplitude $r_{m}$ can be selected
to optimize the nonlocality result. It is known for example, with
$N$ sites and a spin-$1$ system that the optimized state:
\begin{eqnarray}
|\psi\rangle & = & \frac{1}{\sqrt{r^{2}+2}}(|1,-1\rangle^{\otimes N}+r|1,0\rangle^{\otimes N}\nonumber \\
 &  & \ \ \ \ +|1,+1\rangle^{\otimes N}),\label{eq:staterspin1}
\end{eqnarray}
will give improved violation over the maximally entangled state (for
which the amplitudes are uniform) for some Bell inequalities \cite{acin}. }

\textcolor{black}{With the optimization described above, we summarize
}the results explained in Ref. \cite{higherspin steerq} that a growth
of the violation of the nonlocality inequalities for increasing number
$N$ of spin sites is maintained with arbitrary $d$. This is shown
in Figure 2 for qudits $d=2$ and $d=3$ (spin $J=1/2$ and $J=1$),
but means for higher $d$ that one can obtain in principle a violation
of inequalities for arbitrary $d$ by increasing $N$. Thus, quantum
mechanics predicts that at least for some states, increasing contradiction
with separable theories is possible, as the number of sites increases,
even where one has at each site a system of high spin. These results
are consistent with those obtained by other authors \cite{Cabello,multisitequdit,multisitequditson}.

\begin{figure}[h]
\begin{centering}
\includegraphics[scale=0.75]{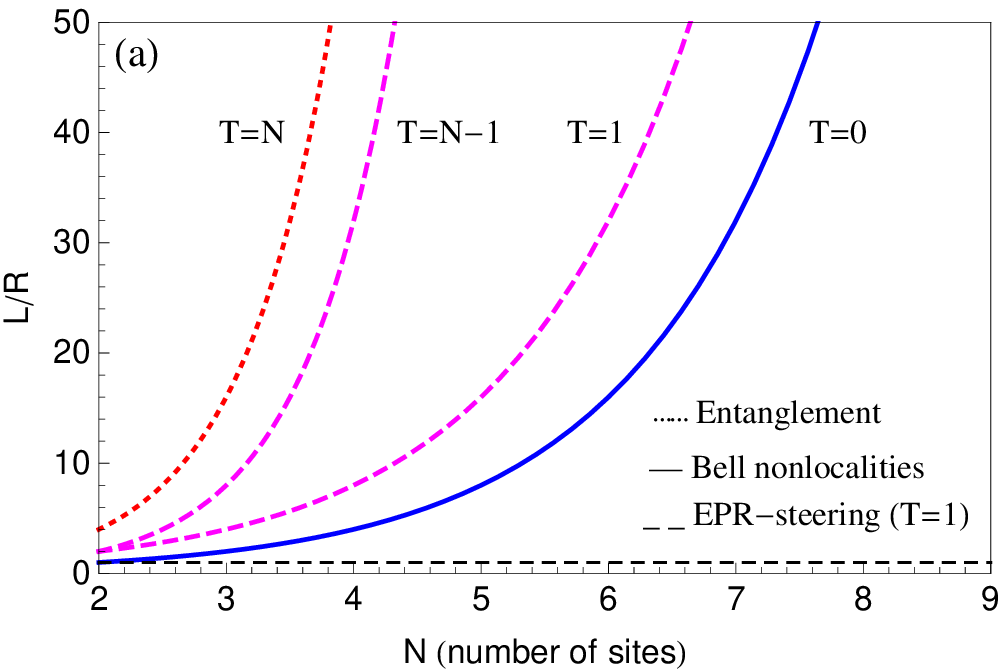}
\par\end{centering}

\begin{centering}
\ \ \includegraphics[scale=0.75]{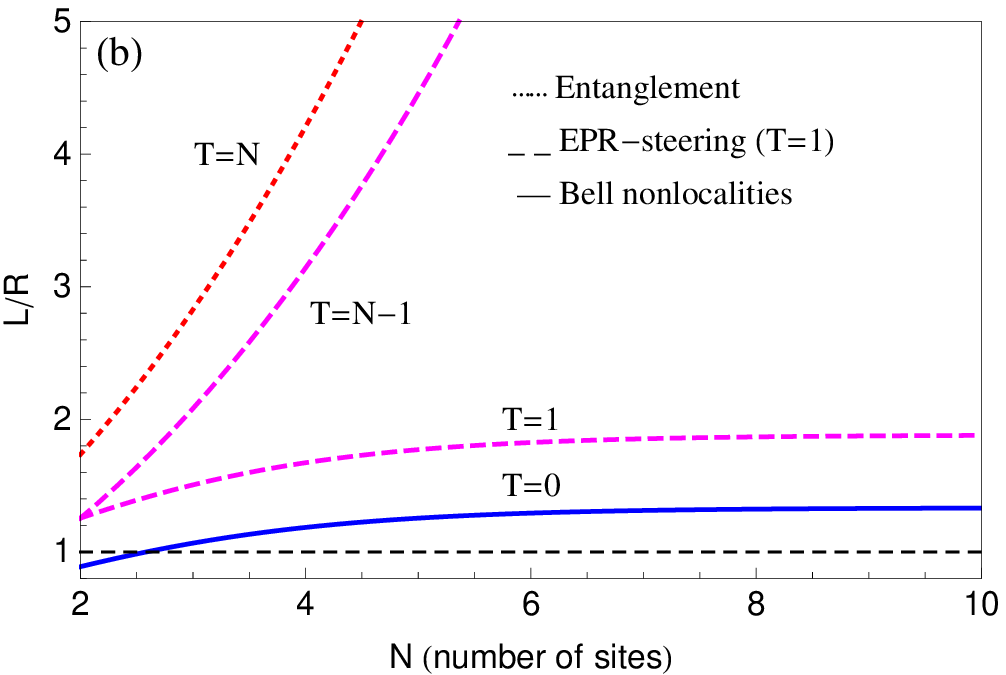}
\par\end{centering}

\caption{\emph{Showing nonlocality to be possible for large numbers $N$ of
spin systems.} The violation of the CFRD Bell ($T=0$), steering ($T=1$)
and entanglement ($T=N$) inequalities of (\ref{eq:ineqcom}), as
measured by the ratio of left side to right side (L/R), for $N$ spin-$1/2$
systems (a), and $N$ spin-$1$ systems (b). Nonlocality is demonstrated
when $L/R>1$.}
\end{figure}

\section{Genuine Multiparticle Nonlocality: qubit example}

Svetlichny \cite{genuine} addressed the following question. How many
particles are \emph{genuinely} entangled? The above nonlocality inequalities
can fail if separability/ locality fails between a single \emph{pair}
of sites. To prove \emph{all} $N$ sites are entangled, or that the
Bell nonlocality is shared between \emph{all} $N$ sites, is a more
challenging task, and one that relates more closely to the question
of multi-particle quantum mechanics. 

To detect genuine nonlocality, one needs to construct different criteria.
For example where $N=3$, to show genuine tripartite entanglement,
we need to exclude that the statistics can be described by bipartite
entanglement i.e., by the models 
\begin{equation}
\rho=\sum_{R}P_{R}\rho_{AB}^{R}\rho_{C}^{R},\,\rho=\sum_{R}P_{R}\rho_{A}^{R}\rho_{BC}^{R},\,\rho=\sum_{R}P_{R}\rho_{B}^{R}\rho_{AC}^{R},\label{eq:genentmodel}
\end{equation}
where $\rho_{IJ}^{R}$ can be \emph{any} density operator for composite
system $I$ and $J$. These models can fail \emph{only} if there is
genuine tripartite entanglement. Thus, to show there is a genuine
tripartite Bell nonlocality, one needs to falsify all models encompassing
bipartite Bell nonlocality, i.e.. 
\begin{equation}
P(x_{\theta},x_{\phi},x_{\vartheta})=\int d\lambda P(\lambda)P_{AB}(x_{\theta},x_{\phi}|\lambda)P_{C}(x_{\vartheta}|\lambda)\label{eq:gennolocmodel}
\end{equation}
and the permutations. In the expansion (\ref{eq:gennolocmodel}),
locality is not assumed between $A$ and $B$, but is assumed between
a composite system $AB$, and $C$. This model allows bipartite entanglement
between $A$ and $B$, but not tripartite entanglement. To test genuine
nonlocality or entanglement, it is therefore useful to consider hybrid
local-nonlocal models. What is a condition for genuine $N$ partite
entanglement? 

Consider again the $N$-qubit system. A recent analysis \cite{ericmulti}
follows Svetlichny \cite{genuine} and Collins \emph{et al. }(CGPRS)
\cite{collspinmol}, to consider a hybrid local-nonlocal model in
which Bell nonlocality \emph{can} exist, but only if shared among
 $k=N-1$ or fewer parties. Separability must then be retained between
any two groups $A$ and $B$ of $ $$k$ and $N-k$ parties respectively,
if $k>N/2$ , and one can write: 
\begin{equation}
\langle\prod_{j=1}^{N}F_{j}^{s_{j}}\rangle=\int_{\lambda}d\lambda P(\lambda)\langle\prod_{j=1}^{k}F_{j}^{s_{j}}\rangle_{A,\lambda}\langle\prod_{j=k+1}^{N}F_{j}^{s_{j}}\rangle_{B,\lambda}.\label{eqn:sepave-1}
\end{equation}
Violation of all such {}``$k$ - nonlocality'' models then implies
the nonlocality to be genuinely {}``$k+1$ partite''. We summarize
Ref. \cite{ericmulti} who use (\ref{eqn:sepave-1}) to consider
consequences of the hybrid model (\ref{eqn:sepave-1}) for the three
different types of nonlocality. Multiplying out $\prod_{j=1}^{N}F_{j}^{s_{j}}=\mathrm{Re}\Pi_{N}+i\mathrm{Im}\Pi_{N}$
reveals recursive relations $\mathrm{Re}\Pi_{N}=\mathrm{Re}\Pi_{N-1}\sigma_{x}^{N}-\mathrm{Im}\Pi_{N-1}\sigma_{y}^{N}$,
$\mathrm{Im}\Pi_{N}=\mathrm{Re}\Pi_{N-1}\sigma_{y}^{N}+\mathrm{Im}\Pi_{N-1}\sigma_{x}^{N}$
which imply algebraic constraints that must hold for all theories
\cite{mermin bellghz}
\begin{eqnarray}
\langle\mathrm{Re}\Pi_{N}\rangle,\langle\mathrm{Im}\Pi_{N}\rangle & \leq & 2^{N-1},\label{eq:alg1}\\
\langle\mathrm{Re}\Pi_{N}\rangle+\langle\mathrm{Im}\Pi_{N}\rangle & \leq & 2^{N}.\label{eq:alg2}
\end{eqnarray}
These recursive relations and the CHSH lemma summarized by Ardehali
\cite{ard} gives the Svetlichny-CGPRS inequality \cite{genuine,collspinmol}
\[
\langle\mathrm{Re}\Pi_{N}\rangle+\langle\mathrm{Im}\Pi_{N}\rangle\leq2^{N-1}
\]
the violation of which confirms genuine $N$ partite Bell-nonlocality.
The quantum prediction maximizes at (\ref{eq:qm2}) to predict violation
by a \emph{constant }amount ($S_{N}=\sqrt{2}$) \cite{ghose-1,ghose2}. 

In order to investigate the other nonlocalities, for example the genuine
multipartite steering, the authors of Ref. \cite{ericmulti} suggest
the hybrid approach of \emph{quantizing} $B$, the group of $N-k$
qubits, but not group $A$. In this case, the extremal points of the
hidden variable product $\langle\Pi_{k}^{A}\rangle_{\lambda}=\langle\prod_{j=1}^{k}F_{j}^{s_{j}}\rangle_{A,\lambda}$
of $A$ is constrained only by the \emph{algebraic} limit (\ref{eq:alg1}),
whereas the product $\langle\Pi_{N-k}^{B}\rangle_{\lambda}\equiv\langle\prod_{j=k+1}^{N}F_{j}^{s_{j}}\rangle_{\lambda}$
for group $B$ is constrained by the \emph{quantum} result (\ref{eq:qm2}).
We note that a criterion for genuine $N$-qubit entanglement is obtained
by constraining \emph{both} $A$ and $B$ to be quantum, leading to
the condition
\begin{equation}
\langle\mathrm{Re}\Pi_{N}\rangle,\langle\mathrm{Im}\Pi_{N}\rangle\leq2^{N-2}
\end{equation}
(as derived in Ref. \cite{tothguhne}), and $\langle\mathrm{Re}\Pi_{N}\rangle+\langle\mathrm{Im}\Pi_{N}\rangle\leq2^{N-3/2}$.
These  are violated by (\ref{eq:qm1}-\ref{eq:qm2}) to confirm genuine
$N$qubit entanglement ($S_{N}=2$).

In short, genuine $N$ particle nonlocality can be confirmed using
MABK Bell inequalities for $N$ qubits, but a higher threshold is
required. The threshold is reached by the quantum prediction of the
GHZ states, but the higher bound implies the level of violation is
\emph{no longer} exponentially increasing with $N$. As a related
consequence, the higher threshold also implies a much higher bound
for efficiency, which makes multi-particle nonlocality difficult to
detect for increasingly larger systems.

\begin{figure}[h]
\begin{centering}
\includegraphics[scale=0.3]{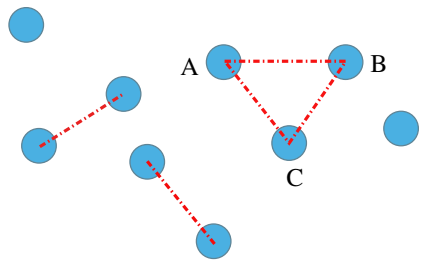} 
\par\end{centering}

\caption{\emph{Genuine nonlocality and entanglement:} Distinguishing entanglement
that may occur between two atoms, and a larger scale entanglement
that necessarily involves more than two atoms. Here is a depiction
of a group of atoms, in which the three atoms $A$, $B$ and $C$
are \emph{genuinely} entangled. We summarize some conditions which
are sufficient to demonstrate an $N$-body nonlocality, whether it
be entanglement, EPR steering, or Bell nonlocality.}
\end{figure}

\section{Investigating Entanglement using Collective measurements: spin squeezing
inequalities}

While detection of individual qubits could be fulfilled in many systems,
the demonstration of a large multi-particle nonlocality would likely
require exceptional detection efficiencies if one is to detect a \emph{genuine}
multi-particle nonlocality for large $N$. We thus review and outline
a complementary approach, which is the measurement of the \emph{collective
}spin of a system.

\subsection{Spin squeezing entanglement criterion}

Consider $N$ identical spin-$J$ particles (Figure 3). One defines
the collective spin operator 
\begin{equation}
J^{X}=\sum_{k=1}^{N}J_{k}^{X}\label{eq:spincoll}
\end{equation}
and similarly a $J^{Y}$ and $J^{Z}$. Entanglement between the spin
$J$ particles can be inferred via measurements of these collective
operators. The concept of spin squeezing was pioneered by Kitagawa
and Ueda \cite{spinsq-1kueda}, and Wineland et al \cite{wineland e}.

To investigate entanglement, we note that for each particle, or quantum
site $k$, the Heisenberg uncertainty relation holds 
\begin{equation}
\Delta J_{k}^{X}\Delta J_{k}^{Y}\geq|\langle J_{k}^{Z}\rangle|/2.\label{eq:hup}
\end{equation}
If the system is fully separable (no entanglement) then 
\begin{equation}
\rho=\sum_{R}P_{R}\rho_{1}^{R}...\rho_{k}^{R}...\rho_{N}^{R}.\label{eq:fulsep}
\end{equation}
For a mixture, the variance is greater than the average of the variances
of the components, which for a product state is the sum of the individual
variances \cite{hofman}. Thus, separability implies 
\begin{equation}
\Delta^{2}J^{X}\geq\sum_{R}P_{R}\sum_{k=1}^{N}\Delta^{2}J_{k}^{X}.\label{eq:varmin}
\end{equation}
The next point to note is that for a fixed dimensionality spin- $J$
system, there is a constraint on the \emph{minimum }value for the
variance of spin. The constraint on the minimum arises because of
the constraint on the \emph{maximum} variance, which for fixed spin
$J$ must be bounded by 
\begin{equation}
\Delta^{2}J^{Y}\leq J^{2}.\label{eq:varmax}
\end{equation}
This implies, by the uncertainty relation, the lower bound on the
minimum variance for a spin J system 
\begin{equation}
\Delta^{2}J^{X}\geq\langle J^{Z}\rangle^{2}/4J^{2}.\label{eq:hupspinsq}
\end{equation}
Then we can prove, using (\ref{eq:varmin}) to get the first line,
\begin{eqnarray}
\Delta^{2}J^{X} & \geq & \frac{1}{4J^{2}}\sum_{k=1}^{N}\sum_{R}P_{R}\langle J_{k}^{Z}\rangle_{R}^{2}\nonumber \\
 & \geq & \frac{1}{4J^{2}}\sum_{k=1}^{N}|\sum_{R}P_{R}\langle J_{k}^{Z}\rangle_{R}|^{2}\nonumber \\
 & = & \frac{1}{4J^{2}}\sum_{k=1}^{N}|\langle J_{k}^{Z}\rangle|^{2}\label{eq:proofspinsq}
\end{eqnarray}
and the Cauchy Schwarz inequality to get the second to last line (use
$(\sum x^{2})(\sum y^{2})\geq|\sum xy|^{2}$ where $x=\sqrt{P_{R}}$
and $y=\sqrt{P_{R}}\langle J_{k}^{Z}\rangle_{R}$). We can rewrite
and use the Cauchy-Schwarz inequality again (this time, $x=1/\sqrt{N}$
and $y=\langle J_{k}^{Z\rangle}\rangle/\sqrt{N}$), to obtain 
\begin{eqnarray}
\Delta^{2}J^{X} & = & \frac{N}{4J^{2}}\sum_{k=1}^{N}\frac{1}{N}|\langle J_{k}^{Z}\rangle|^{2}\nonumber \\
 & \geq & \frac{N}{4J^{2}}|\sum_{k=1}^{N}\frac{1}{N}\langle J_{k}^{Z}\rangle|^{2}\nonumber \\
 & = & \frac{1}{4NJ^{2}}|\langle J^{Z}\rangle|^{2}.\label{eq:proofspinsq-1}
\end{eqnarray}
We can express the result as 
\begin{equation}
y=x^{2}/4J,\label{eq:analyquad}
\end{equation}
where $y=\Delta^{2}J^{X}/J$ and $x=|\langle J^{Z}\rangle|/J$. For
$J=1/2$, we obtain the result that for a fully separable state, 
\begin{equation}
\Delta^{2}J^{X}\geq|\langle J^{Z}\rangle|^{2}/N\label{eq:spinsqentcrit}
\end{equation}
($y=x^{2}/2$). This result for spin $1/2$ was first derived by Sorenson
et al \cite{sorespinsqzoller}, and is referred to as the {}``spin
squeezing criterion'' to detect entanglement. Failure of (\ref{eq:spinsqentcrit})
reflects a reduction in variance (hence {}``squeezing''), and is
confirmation that there is entanglement between at least two particles
(sites). The criterion is often expressed in terms of the parameter
defined by Wineland et al \cite{wineland e}, that is useful measure
of interferometric enhancement, as 
\begin{equation}
\xi=\frac{\sqrt{N}\Delta J^{X}}{|\langle J^{Z}\rangle|}<1.\label{eq:spinsqe}
\end{equation}

\begin{figure}[h]
\begin{centering}
\includegraphics[scale=0.75]{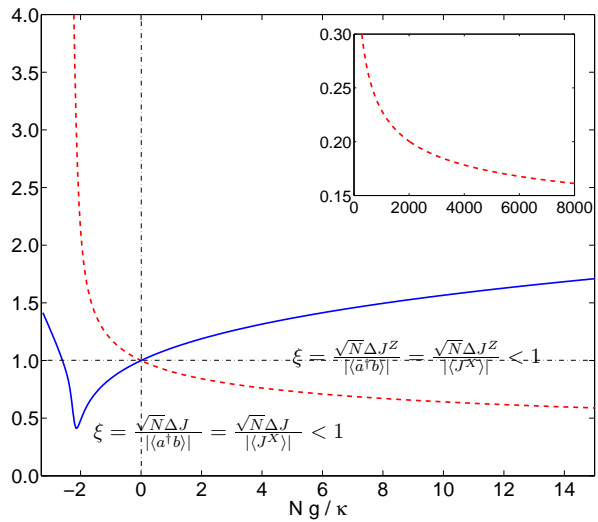} 
\par\end{centering}

\caption{Detecting entanglement within the atoms of a two-component BEC using
the spin squeezing criterion (\ref{eq:spinsqe}) of Sorenson et al
\cite{sorespinsqzoller}. Here, there are $N=100$ atoms in total,
with a fixed intercomponent coupling\textcolor{white}{{} }\textcolor{black}{of
$k/K_{B}=50nK$ and increasing intracomponent interaction} $g$ .
Two states are available to each atom, and the Schwinger spin observables
can be constructed as spin operators, according to (\ref{eq:schwinger}).
The spin squeezing parameter is plotted according to the prediction
of the ground state solution of the Hamiltonian (\ref{hamgs}).\textcolor{red}{{} }}
\end{figure}

The spin squeezing criterion has been used to investigate entanglement
within a group of atoms in a BEC by Esteve et al, Gross et al and
Riedel et al \cite{Germany-spin&entanglement,exp multi,treutnature}.
In fact, spin squeezing is predicted for the ground state of the following
two-mode Hamiltonian
\begin{equation}
H=\kappa(a^{\dagger}b+ab^{\dagger})+\frac{g}{2}[a^{\dagger}a^{\dagger}aa+b^{\dagger}b^{\dagger}bb],\label{hamgs}
\end{equation}
which is a good model for a two-component BEC. Here $\kappa$ denotes
the conversion rate between the two components, and $g$ is a self
interaction term. More details on one method of solution of this Hamiltonian
and some other possible entanglement criteria are given in Ref. \cite{eprbec he}.
To summarize, collective spin operators can be defined in the Schwinger
representation: \textcolor{black}{
\begin{eqnarray}
J^{Z} & = & (a^{\dagger}a-b^{\dagger}b)/2,\nonumber \\
J^{X} & = & (a^{\dagger}b+ab^{\dagger})/2,\nonumber \\
J^{Y} & = & (a^{\dagger}b-ab^{\dagger})/(2i),\nonumber \\
J^{2} & = & \hat{N}(\hat{N}+2)/4,\nonumber \\
\hat{N} & = & a^{\dagger}a+b^{\dagger}b.\label{eq:schwinger}
\end{eqnarray}
}The system is viewed as $N$ atoms, each with two-levels (components)
available to it. For each atom, the spin is defined in terms of boson
operators $J_{i}^{Z}=(a_{i}^{\dagger}a_{i}-b_{i}^{\dagger}b_{i})/2$
where the total number for each atom is $N_{i}=1$, and the outcomes
for $a_{i}^{\dagger}a_{i}$ and $b_{i}^{\dagger}b_{i}$ are $0$ and
$1$. The collective spin defined as $J^{Z}=\sum_{i}J_{i}^{Z}=\sum_{i}(a_{i}^{\dagger}a_{i}-b_{i}^{\dagger}b_{i})/2$
can then be re-expressed in terms of the total occupation number sums
of each level. Figure 4 shows predictions for the variance of the
collective spins $J^{Z}$ or $J^{Y}$, where the mean spin is aligned
along direction $J^{X}$, as a function of ratio $Ng/\kappa$, for
a fixed number of atoms $N$ and a fixed intercomponent coupling.
In nonlinear regimes, indicated by $g\neq0$, we see $\xi<1$ is predicted,
which is sufficient to detect entanglement. Heisenberg relations imply
$\xi\geq1/\sqrt{N}$. 

Sorenson and Molmer \cite{soremol} have evaluated the exact minimum
variance of the spin squeezing for a fixed $J$. Their result for
$J=1/2$ agrees with (\ref{eq:hupspinsq}) and also (\ref{eq:spinsqentcrit}),
but for $J\geq1$ there is a tighter lower bound for the minimum variance,
which can be expressed as 
\begin{equation}
\Delta^{2}J^{X}/J\geq F_{J}(\langle J^{Z}\rangle/J),\label{eq:min functsm}
\end{equation}
where the functions $F_{J}$ are given in Ref. \cite{soremol}. 

The above criteria hold for particles that are effectively indistinguishable.
It is usually of most interest to detect entanglement when the particles
involved are distinguishable, or, even better, causally separated.
We ask how to detect entanglement between spatially-separated or at
least distinguishable groups of spin $J$. We examined this question
in Section III, and considered criteria that were useful for superposition
states with mean zero spin amplitude.

Another method put forward by Sorenson and Molmer (SM) is as follows.
The separability assumption (\ref{eq:fulsep}) will imply \cite{soremol},
\begin{equation}
\Delta^{2}J^{Z}\geq NJF_{J}(\langle J^{X}\rangle/NJ),\label{eq:smolvar}
\end{equation}
where we have for convenience exchanged the notation of $X$ and $Z$
directions (compared to (\ref{eq:min functsm})). The expression applies
when considering $N$ states $\rho_{k}^{R}$ which have a fixed spin
$J$, and could be useful where the mean spin is nonzero.

\subsection{Depth of entanglement and genuine entanglement}

We note from (\ref{eq:min functsm}) that the minimum variance (maximum
spin squeezing) reduces as $J$ increases. Sorenson and Molmer (SM)
showed how this feature can be used to demonstrate that a minimum
number of particles or sites are genuinely entangled \cite{soremol}.
If 
\begin{equation}
\Delta^{2}J^{Z}/NJ<F_{J_{0}}(\langle J^{X}\rangle/NJ),\label{eq:geentpart}
\end{equation}
then we must have $J>J_{0}$ and so a minimum number $N_{0}$ of particles
(where the maximum spin for a block of $N$ atoms is $J=N/2$, we
must have $N_{0}=2J_{0}$ are involved, to allow the higher spin value. 

It will be useful to summarize the proof of this result giving some
detail as follows. Consider a system with the density matrix 
\begin{eqnarray}
\rho & = & \sum_{R}P_{R}\rho^{R}\nonumber \\
\nonumber \\
 & = & \sum_{R}P_{R}\prod_{i=1}^{N_{R}}\rho_{i}^{R}.\label{eq:rho-1}
\end{eqnarray}
We will consider for the sake of simplicity \emph{that the overall
system has a fixed number of atoms $N_{T}$ and a fixed total spin
$J_{tot}$.} The density operator (\ref{eq:rho-1}) describes a system
in a mixture of states $\rho_{R}$, with probability $P_{R}$. For
each possibility $R$, there are $N_{R}$ blocks each with $N_{R,i}$
atoms and a total spin $J_{R,i}$ (note that $J_{R,i}\leq N_{R,i}/2$)
(Figure 5). 

We note that if the maximum number of atoms in each block is $N_{0}$
then the \emph{maximum} spin for the block is $J_{0}=N_{0}/2$. Also,
if the total number of atoms is fixed, at $N_{T}$, then $N_{T}=\sum_{i=1}^{N_{R}}N_{R,i}$,
which implies that each $\rho_{i}^{R}$ has a definite number $N_{R,i}$
, meaning it cannot be a superposition of state of different numbers.
Similarly, for a product state the total spin must be the sum of the
individual spins (as readily verified on checking Clebsch-Gordan coefficients),
which implies that if the total spin is fixed, then each $\rho_{i}^{R}$
has a fixed spin (that is, cannot be in a superposition state of different
spins).

\begin{figure}[h]
\begin{centering}
\includegraphics[scale=0.3]{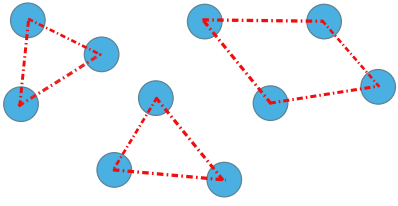} 
\par\end{centering}

\caption{\emph{Genuine multi-particle entanglement:} The SM approach is to
consider that the $N_{T}$ atoms may be described by product states
of blocks of up to $N_{0}$ entangled atoms.\textcolor{red}{{} }\textcolor{black}{Here
$N_{0}=4$.}}
\end{figure}

Using again that the variance of the mixture cannot be less than the
average of its components, and that the variance of the product state
$\rho^{R}$ is the sum of the variances $(\Delta^{2}J^{Z})_{R,i}$
of each factor state $\rho_{i}^{R}$, we apply (\ref{eq:min functsm}),
that the variance has a lower bound determined by the spin $J_{R,i}$.
Thus we can write:

\begin{eqnarray}
\Delta^{2}J^{Z} & \geq & \sum_{R}P_{R}\sum_{i=1}^{N_{R}}(\Delta^{2}J^{Z})_{R,i}\nonumber \\
 & \geq & \sum_{R}P_{R}\sum_{i=1}^{N_{R}}J_{R,i}F_{J_{R,i}}(|\langle J^{X}\rangle|_{R,i}/J_{R,i}).
\end{eqnarray}
\textcolor{black}{Now we can use the fact that the curves $F_{J}$
are nested to form a decreasing sequence at each value of their domain,
as $J$ increases, as explained by Sorenson and Molmer. We then apply
the steps based on the SM proof (lines (6) -(8) of their paper), which
uses convexity of the functions $F_{J}$. We cannot exclude that the
total spin of a block can be zero, $J_{R,i}=0$, for which $(\Delta^{2}J^{Z})_{R,i}\geq0$,
but such blocks do not contribute to the summation and can be formally
excluded. We define the total spin $\sum_{i=1}^{N_{R}}J_{R,i}=J_{tot}^{R}$
for each $\rho^{R}$ but note that for fixed total spin this is equal
to $J_{tot}$, and we also note that $J_{tot}^{R}\leq J_{0}$. In
the later steps below, we define the total spin as $J_{tot}=\sum_{R}P_{R}J_{tot}^{R}$
and the collective spin operator $J^{Z}$.}

\begin{eqnarray}
\Delta^{2}J^{Z} & \geq & \sum_{R}P_{R}\sum_{i=1}^{N_{R}}J_{R,i}F_{J_{0}}(\langle J^{X}\rangle_{R,i}/J_{R,i})\nonumber \\
 & = & \sum_{R}P_{R}J_{tot}^{R}\sum_{i=1}^{N_{R}}\frac{J_{R,i}}{J_{tot}^{R}}F_{J_{0}}(\langle J^{X}\rangle_{R,i}/J_{R,i})\nonumber \\
 & \geq & \sum_{R}P_{R}J_{tot}^{R}F_{J_{0}}(\frac{\sum_{i=1}^{N_{R}}\langle J^{X}\rangle_{R,i}}{J_{tot}^{R}})\nonumber \\
 & = & J_{tot}\sum_{R}P_{R}\frac{J_{tot}^{R}}{J_{tot}}F_{J_{0}}(\frac{\sum_{i=1}^{N_{R}}\langle J^{X}\rangle_{R,i}}{J_{tot}^{R}})\nonumber \\
 & \geq & J_{tot}F_{J_{0}}(\sum_{R}P_{R}\frac{1}{J_{tot}}\sum_{i=1}^{N_{R}}\langle J^{X}\rangle{}_{R,i})\nonumber \\
 & \geq & J_{tot}F_{J_{0}}(\frac{1}{J_{tot}}\sum_{R}P_{R}\sum_{i=1}^{N_{R}}\langle J^{X}\rangle{}_{R,i})\nonumber \\
 & = & J_{tot}F_{J_{0}}(\langle J^{X}\rangle/J_{tot}).\label{eq:proof2-1}
\end{eqnarray}
The total spin $J_{tot}$ is maximum at $J_{tot}=N/2$ where $N$
is total number of atoms over both systems, but is assumed measurable.
Thus, if the maximum number of atoms in each block does not exceed
$N_{0}$, then the inequality (\ref{eq:proof2-1}) must always hold.
The violation of (\ref{eq:proof2-1}) is a demonstration of a group
of atoms that are genuinely entangled \cite{soremol}.

\begin{figure}[h]
\begin{centering}
$\ \ \ \ \ \ $\includegraphics[scale=0.75]{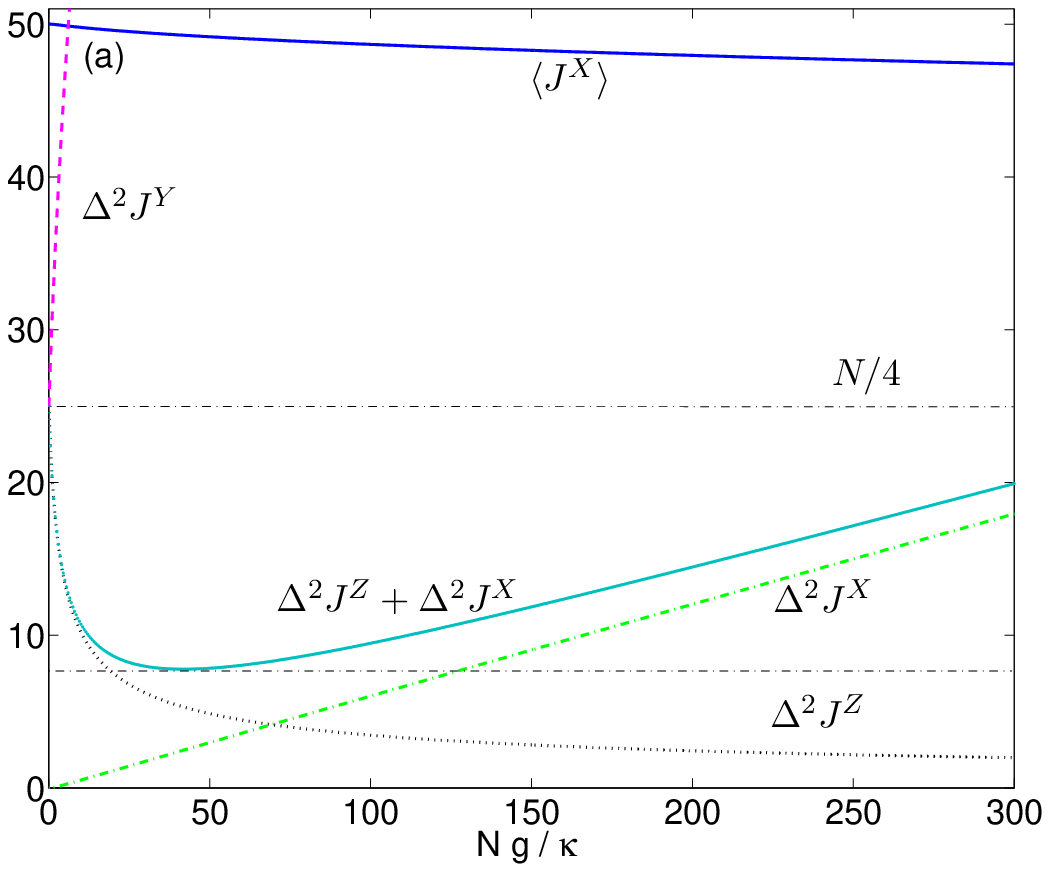}
\par\end{centering}
\smallskip{}
\smallskip{}
\begin{centering}
\includegraphics[scale=0.75]{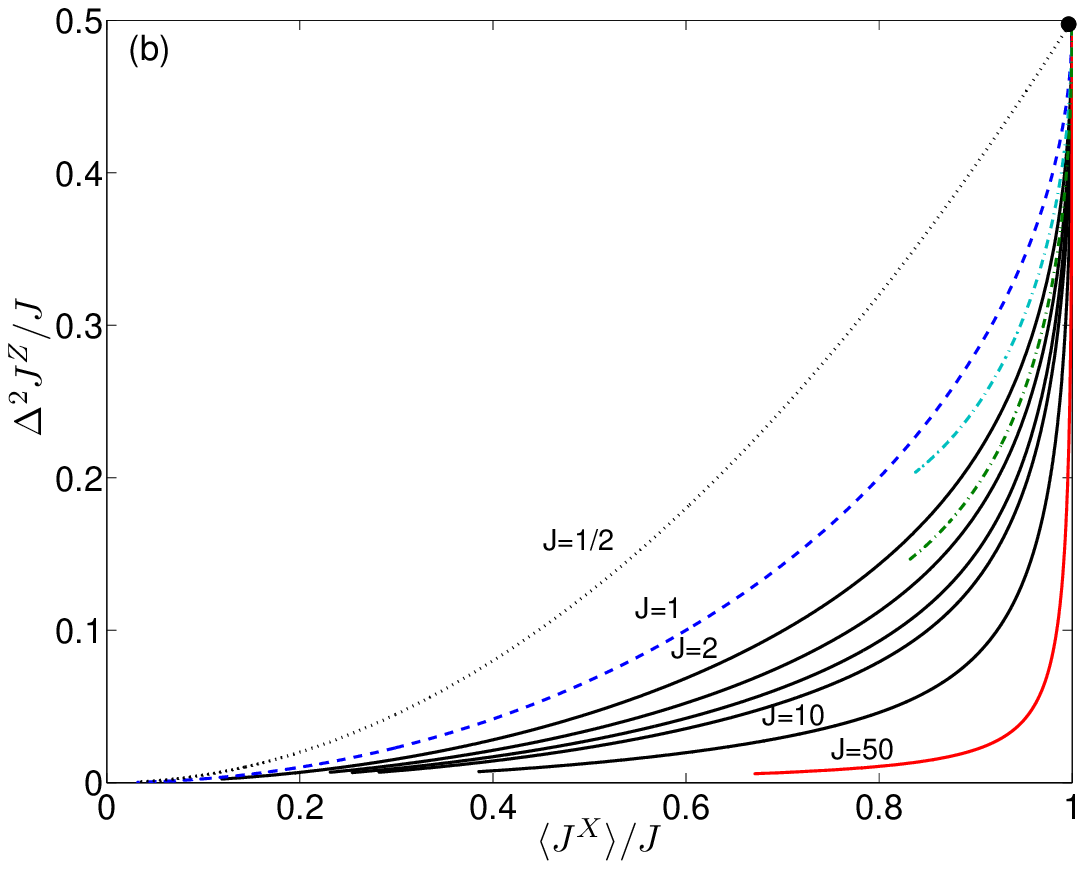}
\par\end{centering}

\caption{Detecting multi-particle entanglement in ground state of a two-component
BEC, as modeled by ({\ref{hamgs}}). (a) The predictions
according to {(\ref{hamgs}}) for the spin moments
for 100 atoms ($N=100$), where there is a fixed conversion rate of
$k/K_{B}=50nK$, but an increasing intrawell interaction $g$. (b) The corresponding prediction for the ratio
$\Delta^{2}J^{Z}/J$ as a function of normalized mean spin amplitude
$\langle J^{X}\rangle/J$ for different $N$, so that the SM inequality
(\ref{eq:proof2-1}) can be tested.}

\end{figure}

The predictions of the model (\ref{hamgs}) are given in Figure 6,
for a range of values of $N$ (the total number of atoms). \textcolor{black}{In
each case, there is a constant total spin, $J\equiv J_{tot}$, given
by $\langle(J^{X})^{2}+(J^{Y})^{2}+(J^{Z})^{2}\rangle=J(J+1)$ where
$J=N/2$. We keep $N$ and the interwell coupling $\kappa$ fixed,
and note that the variance of $J^{Z}$ decreases with increasing $g$,
while the variance in $J^{X}$ increases. Evaluation of the normalized
quantities of the SM inequality (\ref{eq:proof2-1}) are given in
the second plot of Figure 6. Comparing with the functions $F_{J_{0}}$
reveals the prediction of a full $N$ particle entanglement, where
$N$ is an integer value. }
      
We note this treatment does not itself test nonlocality, or even the
quantum separability models (\ref{eq:sepN}-\ref{eq:sepent}) because
measurements are not taken at distinct locations. However, it can
reveal, \emph{within a quantum framework}, an underlying entanglement,
of the type that could give nonlocality if the individual spins could
be measured at different locations. The great advantage however of
the collective criteria is the reduced sensitivity to efficiency,
since it is no longer necessary to measure the spin at each site.
The depth of spin squeezing has been used recently and reported at
this conference to infer blocks of entangled atoms in BEC condensates
\cite{exp multi,treutnature}.

To test nonlocality between sites, the criteria need will involve
measurements made at the different spatial locations. How to detect
entanglement between two-modes using spin operators \cite{hillzub,schvogent,spinsq,toth2009,spinsqkorb},
and how to detect a true Einstein-Podolsky-Rosen (EPR) entanglement
\cite{epr rev ,reidepr,eprbohmparadox,sumuncerduan,spinprodg,Kdechoum,cavalreiduncer}
in BEC \cite{murraybecepr,eprbecbar,eprbec he} are the topics of
much current interest.

\subsection{EPR steering nonlocality with atoms }

An interesting question is whether one derive criteria, involving
collective operators, to determine whether there are stronger underlying
nonlocalities. How can we infer whether the one group of atoms $A$
can {}``steer'' a second group $B$, as shown in schematic form
in Figure 7? This would confirm an EPR paradox between the two groups,
that the correlations imply inconsistency between Local Realism (LR)
and the completeness of quantum mechanics. This is an interesting
task since very little experimental work has been done on confirming
EPR paradoxes between even single atoms. Steering paradoxes between
groups of atoms raise even more fundamental questions about mesoscopic
quantum mechanics.

\begin{figure}[h]
\begin{centering}
\includegraphics[scale=0.4]{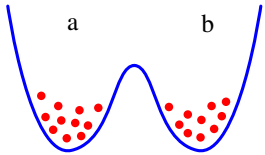} 
\par\end{centering}

\caption{Is {}``EPR steering'' of one group of atoms by another group possible?
How can we detect such steering? }
\end{figure}

As an example, we thus consider the following. EPR steering is demonstrated
between $N$ sites when the LHS model (\ref{eq:bell-1}) fails with
$T=1$ fails. The system (which we will call $B$) at the one site
corresponding to $T=1$ is described by a local quantum state LQS,
which means it is constrained by the uncertainty principle. All other
groups are described by a Local Hidden Variable Theory (LHV), and
thus are constrained to have only a non-negative variance. For this
first group (only) there is the SM minimum variance (implied by quantum
mechanics): 
\begin{equation}
\Delta^{2}J_{B}^{X}\geq J_{B}F_{J}(\langle J^{Z}\rangle/J_{B})\label{eq:smolvar-1}
\end{equation}
Hence, with this assumption, we follow the approach of Section V.
B, to write (where we assume the maximum spin of the steered group
$B$ is $J_{0}$)

\begin{eqnarray}
\Delta^{2}J^{X} & \geq & \sum_{R}P_{R}\{J_{R,B}F_{J_{0}}(\langle J^{Z}\rangle_{R,B}/J_{R,B})\}\nonumber \\
 & = & \sum_{R}P_{R}J_{tot,}^{R}\frac{J_{R,B}}{J_{tot}^{R}}F_{J_{0}}(\langle J^{Z}\rangle_{R,B}/J_{R,B})\nonumber \\
 & \geq & \sum_{R}P_{R}J_{tot}^{R}F_{J_{0}}(\frac{\langle J^{Z}\rangle_{R,B}}{J_{tot}^{R}})\nonumber \\
 & = & J_{tot}\sum_{R}P_{R}\frac{J_{tot}^{R}}{J_{tot}}F_{J_{0}}(\frac{\sum_{i=1}^{N_{R}}\langle J^{Z}\rangle_{R,i}}{J_{tot}^{R}})\nonumber \\
 & \geq & J_{tot}F_{J_{0}}(\sum_{R}P_{R}\frac{1}{J_{tot}}\langle J^{Z}\rangle{}_{R,B})\nonumber \\
 & \geq & J_{tot}F_{J_{0}}(\frac{1}{J_{tot}}\sum_{R}P_{R}\langle J^{Z}\rangle{}_{R,B})\nonumber \\
 & = & J_{tot}F_{J_{0}}(\langle J_{B}^{Z}\rangle/J_{tot})\label{eq:proof2-1-1}
\end{eqnarray}
If the inequality is violated, a {}``steering'' between the two
groups is confirmed possible: group $A$ {}``steers'' group $B$.
In this case, the spins of spatially separated systems $B$ would
need to be measured, and potential such {}``EPR'' systems have been
proposed, with a view to this sort of experiment in the future.

\section{Conclusion}

We have examined a strategy for testing multi-particle nonlocality,
by first defining three distinct levels of nonlocality: (1) entanglement,
(2) EPR paradox/ steering, or (3) failure of local hidden variable
(LHV) theories (which we call Bell's nonlocality). We next focused
on two types of earlier studies that yielded information about nonlocality
in systems of more than two particles. 

The first study originated with Greenberger, Horne and Zeilinger (GHZ)
and considers $N$ spatially separated in $1/2$ particles, on which
individual spin measurements are made. The study revealed that nonlocality
involving $N$ spatially separated (spin $1/2$) particles can be
more extreme. Mermin showed that the deviation of the quantum prediction
from the classical LHV boundaries can grow exponentially with $N$
for this scenario. Here we have summarized some recent results by
us that reveal similar features for entanglement and EPR steering
nonlocalities. Inequalities are presented that enable detection of
these nonlocalities in this multipartite scenario, for certain correlated
quantum states. The results are also applicable to $N$ spin $J$
particles (or systems), and thus reveal nonlocality can survive for
$N$ systems even where these systems have a higher dimensionality. 

We then examined the meaning of {}``multi-particle nonlocality'',
in the sense originated by Svetlichny, that there is an {}``$N-$body''
nonlocality, necessarily shared among \emph{all} $N$systems. For
example, three-particle entanglement is defined as an entanglement
that cannot be modeled using two-particle entangled or separable states
only. Such entanglement, generalized to $N$ parties, is called genuine
$N$ partite entanglement. We present some recent inequalities that
detect such genuine nonlocality for the GHZ/ Mermin scenario of $N$
spin $1/2$ particles, and show a higher threshold is required that
will imply a much greater sensitivity to inefficiencies $\eta$. In
other words, the depth of violation of the Bell or nonlocality inequalities
determines the level of \emph{genuine} multi-particle nonlocality.

This led to the final focus of the paper, which examined criteria
that employ collective spin measurements.\textcolor{red}{{} }\textcolor{black}{For
example,}\textcolor{red}{{} }the spin squeezing entanglement criterion
of Sorenson et al enables entanglement to be confirmed between $N$
spin $1/2$ particles, based on a reduction in the overall variance
({}``squeezing'') of a single collective spin component. The criterion
works because of the finite dimensionality of the spin Hilbert space,
which means only higher spin systems $-$ as can be formed from entangled
spin $1/2$ states $-$ can have larger variances in one spin component,
and hence smaller variances in the other. As shown by Sorenson and
Molmer, even greater squeezing of the spin variances will imply larger
entanglement, between more particles. Hence the depth of spin squeezing,
as with the depth of Bell violations in the GHZ Mermin example above,
will imply genuine entanglement between a minimum number of particles.
This result has recently been used to detect experimental multi-particle
entanglement in BEC systems. We present a model of the ground state
BEC for the two component system, calculating the extent of such multi-particle
squeezing. 

We make the final point that, while collective spin measurements are
useful in detecting multi-particle entanglement and overcoming problems
that are encountered with detection inefficiencies, the method does
not address tests of nonlocality unless the measured systems can be
at least in principle spatially separated. This provides motivation
for studies of entanglement and EPR steering between groups of atoms
in spatially distinct environments.
\begin{acknowledgments}

\end{acknowledgments}
We wish to thank the Humboldt Foundation, Heidelberg University, and
the Australian Research Council for funding via AQUAO COE and Discovery
grants, as well as useful discussions with Markus Oberthaler, Philip
Treutlein, and Andrei Sidorov.

\end{document}